\documentstyle[12pt]{article}
\def\bsg{\ $b \rightarrow s \gamma$\ }
\def\bsm{\ $B_s \rightarrow \mu^+ \mu^-$\, }
\def\beq{\begin{equation}}
\def\eeq{\end{equation}}

\begin{document}
\begin{titlepage}
\begin{center}
May, 1993     \hfill      MIT-CTP-2211 \\
 \hfill     LBL-34110 \\
                    \hfill       UCB-PTH-93/17 \\
\vskip 0.2 in
{\large \bf $b \rightarrow s \gamma$ and $B_s \rightarrow \mu^+ \mu^-$ in
Extended Technicolor Models
}

\vskip .2 in
       {\bf Lisa Randall}$^{(1)}$\footnotetext{$(1)~$ Supported
in part by funds provided by the U.S.
Department of Energy (DOE) under contract \#DE-AC02-76ER03069 and in
part by the Texas National Research Laboratory Commission
under grant \#RGFY92C6.\hfill\break
National
Science Foundation Young Investigator Award.\hfill\break
Alred P.~Sloan
Foundation Research Fellowship.\hfill\break
Department of Energy Outstanding Junior
Investigator Award.}
\vskip 0.3cm
{\it Massachusetts Institute of Technology\\
Cambridge, MA 02139\\
}
         {\it and \\}
{\bf Raman Sundrum}$^{(2)}$\footnotetext{$(2)~$ Supported in part by the
 Director, Office
of Energy Research, Office of High Energy and Nuclear Physics, Division of High
Energy Physics of the U.S. Department of Energy under contract
DE-AC03-76SF00098 and in part by the National Science Foundation under grant
PHY90-21139.}
        \vskip 0.3 cm
       {\it Department of Physics\\
          University of California, Berkeley, CA 94720\\
          and\\
          Lawrence Berkeley Laboratory \\
          1 Cyclotron Road, Berkeley, CA 94720 \\   }
        \vskip 0.3 cm

\begin{abstract}

The rates of the rare flavor-changing processes, $b \rightarrow s \gamma$
and $B_s \rightarrow \mu^+ \mu^-$ are estimated in
extended technicolor models with
and without a GIM mechanism. We find the $b \rightarrow s \gamma$ rate in
ETC models with a GIM mechanism to be at most slightly larger than
the standard model rate, whereas there is no significant extra
model-independent contribution in other  ETC scenarios.
 In the case of $B_s \rightarrow \mu^+ \mu^-$, ETC models with a GIM
mechanism can yield a rate up to two orders of magnitude
bigger than that of the standard model, whereas generic ETC
scenarios are likely to give a rate
which is about an order of magnitude bigger than that of the
standard model.

\end{abstract}
\end{center}

\end{titlepage}

\section{Introduction}

Rare flavor-changing processes involving the third generation quarks are
sensitive probes of the flavor sector because of the large couplings of the top
quark. Two good candidates for study are \bsg and \bsm.
The process \bsg has recently been observed at a rate
compatible with that of the standard model \cite{bgam}. Models
which give a contribution significantly larger than that
expected in the standard model are therefore ruled out.
 The process \bsm may be studied very well in
the future \cite{bmu}. Much attention has been devoted to the study
of both these processes within the context of two-Higgs models
and supersymmetry \cite{2hig,sbmu}.
In this paper we examine these processes  in
extended technicolor (ETC) theories.

We will consider two types of ETC models. In the
first class of models, which we refer to as ``traditional" ETC,
we assume the  existence of the minimal features needed
to describe the ETC origins of the third generation quark masses.
We will estimate  contributions to \bsg and \bsm arising from
this structure. Since no realistic models in this class are known our estimates
should be taken as lower bounds (barring large cancellations)
on the nonstandard contributions, since it
 may be
that a realistic model will require extra interactions which also contribute to
the processes of interest.
We will only consider models with a single techni-doublet breaking electroweak
symmetry. Larger technicolor sectors are disfavored by electroweak
$S$-parameter constraints \cite{s}.

The second type of model we consider are those
which incorporate a techni-GIM mechanism \cite{gim,tgim,models,model2}.
 Recently, it has been shown \cite{models,model2}
that realistic technicolor models can be built
incorporating a techni-GIM mechanism which protects them from large
flavor-changing neutral currents (FCNC's). Such models contain many
approximately degenerate ETC gauge bosons, with masses at most a few TeV in
order to obtain the heavy top quark mass.
This is to be contrasted with traditional ETC models where the
ETC gauge bosons are highly non-degenerate, which is related to the
difficulties such models have in adequately
 suppressing FCNC's involving the first two generations \cite{fcnc}.

The large number of TeV-scale ETC particles in techni-GIM models can mediate
important low-energy effects. In particular, it is quite
natural in such models to
have quarks and leptons in a common ETC group, so that ETC gauge-boson
exchange can
induce four-fermion interactions involving both quarks and leptons which can
greatly enhance the \bsm rate.
 In a ``traditional" ETC scenario, it
is  expected that the scale for such four-fermion interactions
is much larger, determined by the ETC scale of the leptons.

Our results are as follows. We find in both classes
of models that the generic ETC rate for \bsg is essentially the same as in  the
standard model. In techni-GIM models with strongly interacting ETC
dynamics it possible for the rate to be at most  slightly larger than in the
standard
model. On the other hand, the process \bsm is likely to be a good test of ETC
in either class of model, but will be particularly
sensitive in the models  of quarks and leptons which
incorporate a techni-GIM mechanism. Generically, ETC models of either class
predict a rate for \bsm which is an order of magnitude larger than the standard
model rate (for fixed top quark mass), while the simplest techni-GIM models,
which have leptons and quarks in the same ETC multiplets, predict a rate
two orders of magnitude larger than in the standard model.

We are restricting our attention to
models in which neither
 composite \cite{comp} nor fundamental \cite{fund} scalars
are involved in the quark mass generation. Their presence could substantially
weaken the  nonstandard contributions
to the rare decays we find in this paper. However, the estimates we
make should not be weakened much by  walking technicolor
dynamics \cite{walk} because walking affects the dynamics far above the weak
scale while third generation ETC scales are necessarily low.

Some of our calculations require technicolor strong interaction matrix
elements. We  estimate these using naive dimensional analysis (NDA) \cite{nda}.

In section 2, we relate some parameters of
the effective technicolor lagrangian to the fermion masses and mixings,
in particular the top quark mass.
 In section 3, we estimate the rate for $B_s \rightarrow \mu^+
 \mu^-$ in ETC theories with and without a GIM mechanism.
In section 4, we do the same for  the process $b \rightarrow s \gamma$.
 In section 5, we provide our conclusions.

\section{Fermion masses and ETC interactions}

In either of the  ETC scenarios we consider, exchange of ETC gauge bosons
 induces interactions at the TeV scale of the form
\begin{eqnarray}
{\cal L}_{mass} = {1 \over f^2}
 (\overline{\psi}_L^i \gamma_{\mu} T_L)(\overline{U}_R
\gamma^{\mu} u_R^j) Y_u^{ij} +
{1 \over f^2} (\overline{\psi}_L^i \gamma_{\mu} T_L)(\overline{D}_R
\gamma^{\mu} d_R^j) Y_d^{ij} +  \rm h.c.,
\end{eqnarray}
necessary for giving quark masses upon electroweak symmetry breaking.
Here,  $T_L$ is the techni-doublet while $U_R, D_R$ are the right-handed
partners and $\psi_L^i$ is the $i$-th generation quark doublet and $u_R^i,
{}~d_R^i$ are its right-handed partners. The matrices $Y_{u(d)}$ play a role
here
similar to the standard model Yukawa couplings. They parameterize the
breaking of the  $SU(3)_L \times
SU(3)_u \times SU(3)_d$ symmetry under which the three generations of
left-handed quark doublets, right-handed up-type quarks and right-handed
down-type quarks transform. This symmetry must be broken to allow for quark
masses. We will work with the normalization of the
$Y$'s which follows most naturally in techni-GIM models, namely their largest
element equals one, $Y_u^{33} = 1$.

An important difference between techni-GIM models and traditional ETC models
is that in techni-GIM models the ETC physics ensures that
the $SU(3)_L \times SU(3)_u \times SU(3)_d$ symmetry of the full effective
lagrangian at the TeV scale (resulting from integrating out heavier ETC
physics) is broken {\it only} by the parameters
$Y_u, Y_d$. In traditional models the full effective lagrangian contains a more
general form of flavor symmetry breaking. It is the restricted form of the
flavor symmetry breaking in techni-GIM models that provides the GIM suppression
of FCNC's.

The two classes of models also differ in the way ETC physics induces
eq. (1). In techni-GIM models, the ETC gauge bosons are approximately
degenerate in mass, and the ETC scale $f$ is given by
\begin{equation}
{1 \over f^2} = {g_{ETC}^2 \over m_{ETC}^2},
\end{equation}
where $g_{ETC}$ is the coupling of the ETC gauge bosons and $m_{ETC}$ is their
mass.  The slight non-degeneracy of the ETC gauge bosons is responsible for the
breaking of the $SU(3)_L \times SU(3)_u \times SU(3)_d$ flavor symmetry. An
individual quark mass eigenstate gets contributions to its mass from the
exchange of several different ETC gauge boson mass eigenstates. This is to be
contrasted with traditional ETC models where the high level of quark
non-degeneracy is reflected in a high level of non-degeneracy in the ETC gauge
boson spectrum, and a quark mass eigenstate gets contributions to its mass from
essentially a single ETC gauge boson mass eigenstate. The fact that top
quark mass production is associated with a much lower ETC scale than bottom
quark mass production can give rise to unacceptable
violation of the $\rho$ relation \cite{zfc}.
In techni-GIM ETC models this is not a problem because
 the two ETC scales can be  taken
 nearly equal \cite{models} while still allowing a large top-bottom
 splitting.

Upon technifermion condensation, quark mass matrices emerge,
\begin{eqnarray}
M_{u(d)} \approx {4 \pi v^3 Y_{u(d)} \over f^2},
\end{eqnarray}
where $v$, the technifermion decay constant is taken to be the weak scale,
$246$ GeV, in order to ensure the correct $W, Z$ masses. In particular, since
$Y_u^{33} = 1$ we have,
\beq
m_t \approx {4 \pi v^3 \over f^2}.
\eeq
We will use this relation in our estimates to eliminate the ETC scale $f$ in
favor of $m_t$.

\section{\bsm}

In both classes of models the generic ETC contribution to \bsm is mediated by
$Z$ exchange, with ETC-induced flavor-changing couplings of the $Z$.
 We begin with traditional ETC models. ETC gauge boson
exchange can induce the operator
\begin{equation}
\xi {g_{ETC}^2 \over m_{ETC}^2} ~ (\overline{\psi}_L \gamma_{\mu} T_L)
(\overline{T}_L \gamma^{\mu} \psi_L),
\end{equation}
where $\psi_L$ is the doublet containing the top quark. The ETC scale appearing
here is therefore expected to be that associated with top quark mass
production, namely $\xi$ is a model dependent parameter
 of order one and $g_{ETC}^2/m_{ETC}^2 \approx 1/f^2$, with $f$ as in eq. (4).
 Proceeding as in refs. \cite{zbb,zbb2,zfc}, this translates
into a $Z$ coupling,
 \beq\label{op}
 \xi {m_t \over 16 \pi v} \overline{\psi_L}
 \left({e \over \sin \theta \cos \theta} Z\!\! \!\!\slash \tau^3
 \right)\psi_L,
 \eeq
where $\theta$ is the weak mixing angle. There is no technicolor strong
interaction
uncertainty in this computation because the technifermion pair
coupling to the $Z$ is the same as the current involved in the Higgs mechanism
for electroweak symmetry breaking.
Because the bottom quark field in $\psi_L$  is not the mass eigenstate,
  one expects flavor-changing effects
of order $m_t/16 \pi v$ times a mixing angle between
the second and third generation.  (There
are of course also flavor diagonal modifications of the $Z$
vertex \cite{zbb}. However, traditional ETC models in which these nonstandard
effects are large
 may also unacceptably violate the well constrained $\rho$ relation
 \cite{zfc}.)

A similar modification of the $Z$ vertex  occurs
in models with a GIM mechanism.  In this case,
the effective lagrangian contains the operator
\begin{equation}
{\xi_1 \over f^2} ~(\overline{\psi}^i_L \gamma_{\mu} T_L)
(Y_u Y_u^{\dag})_{ij} (\overline{T}_L \gamma_{\mu} \psi_L^j),
\end{equation}
where $\xi_1$ is an order one, model-dependent coefficient.
Proceeding as before, the induced coupling to the $Z$ is
\begin{equation}
 {\xi_1 v^2 \over 4 f^2}~ \frac{e}{c_{\theta} s_{\theta}}~
\overline{\psi}_L Z\!\! \!\!\slash \tau_3 Y_u Y_u^{\dag} \psi_L.
\end{equation}
This term contains the large flavor-changing vertex
\begin{equation}
-{\xi_1 m_t \over 16 \pi v}~ \frac{e}{c_{\theta} s_{\theta}} ~
\overline{b}_L Z\!\! \!\!\slash ~Y_u^{33} Y_u^{\dag 32} s_L.
\end{equation}
For numerical estimates we will take $Y_u^{23} \sim V_{ts}$.

Therefore, in both classes of ETC models, $Z$ exchange yields the operator
\beq
\overline{s}_L \gamma^\mu  b_L \overline{\mu} \gamma_\mu \gamma_5 \mu
\eeq
with coefficient
\beq
C_{ETC} \sim {m_t V_{ts} \over 16 \pi v^3}.
\eeq
For comparison, in the standard model, the same operator is induced
at the one loop level \cite{sbmu}, with coefficient,
\beq
C_{SM}={g^2 (B({m_t \over m_W})^2)-C({m_t \over m_W})^2) V_{ts} \over
8 \pi^2 v^2},
\eeq
where  $g$ is the SU(2) gauge coupling and
the functions $B(x)$ and $C(x)$ are order one over the top mass range and are
given explicitly in ref. \cite{sbmu}.

We normalize to the semileptonic
decay width given by
\beq
\Gamma( b \to c e^- \overline{\nu}_e)={G_F^2 m_b^5 \over 192 \pi^3}
G\left({m_c \over m_b}\right) |V_{bc}|^2,
\eeq
where $G(m_c/m_b) \approx {1 \over 2}$ takes into account the large
charm quark mass.
The rate for $B_s \to \mu^+ \mu^-$ is
\beq
{C^2 m_B m_\mu^2 f_B^2 \over 8 \pi}
\eeq
where $C$ is the coefficient of the operator given above.
Therefore the rate, relative to the semileptonic rate
to electron final state given above, is
\beq
{\Gamma(B_s \rightarrow \mu^+ \mu^-)
\over \Gamma( b \to c e^- \overline{\nu}_e)}
\approx 2 \cdot 10^{-7} \left({m_t \over v} \right)^2
\left({f_B \over 200~ {\rm MeV}} \right)^2,
\eeq
 where we have taken $Y^{23}_u \approx V_{ts} \approx V_{bc}$.

There can also be direct contributions in
technicolor coming from operators in the effective lagrangian,
\begin{equation}
{ \xi_2 \over f^2} ~(\overline{\psi}_L \gamma_{\mu}
Y_u Y_u^{\dag}  \psi_L) (\overline{l}_L \gamma^{\mu} l_L),
\end{equation}
where $l_L$ are the left-handed lepton doublets.
The size of $\xi_2$ is model dependent. (In traditional ETC, we might expect
a similar interaction to be weak, linked to a large leptonic ETC scale,
in which case it can be neglected.)
In techni-GIM models like those of refs. \cite{models,model2} where we expect
all ETC gauge boson masses to be approximately equal,
$\xi_2$ should be of order unity.
In this case the decay rate is larger than the $Z$-exchange contribution,
\begin{equation}
{\Gamma(B_s \rightarrow \mu^+ \mu^-)
\over \Gamma( b \to c e^- \overline{\nu}_e)}
\sim 1.5  \cdot 10^{-6}
 \left({m_t \over v} \right)^2 \left({f_B \over 200~ {\rm MeV}} \right)^2.
\end{equation}

The relative decay rates in ETC and in the standard model
 are plotted against $m_t$
in fig. (1). As can be seen, for a potentially attainable $10^{-6}$--$10^{-7}$
sensitivity
 one may be able to just measure  the ETC induced rare decays if
only $Z$-exchange contributes,
while the prospects are good for doing so in  models in which there is a
direct ETC contribution.

In ref. \cite{zfc}, various other flavor violating effects related to the top
quark were considered.
It was shown that the strongest constraint came from $B$--$\overline{B}$
mixing, implying a favored region of KM angles which was
more restrictive than that of the standard model.   Here we
have only used mixing between the second and third generation,
which is directly determined from the $B$ lifetime, so
the altered parameter region is not relevant to the two decays
considered here (although it would be relevant to other decays as discussed
in \cite{zfc}).

\section{$b \rightarrow s \gamma$}

It is important to notice that it is only the $Z$,
not the photon vertex
which is corrected at leading order. Gauge invariance
guarantees there is no renormalization of the dimension-four photon coupling.
Moreover, one should recognize that the reason the
modification of the $Z$ coupling
was so large was that it was a correction to a low dimension operator.
In particular, if one estimates (despite it being
exactly determined), using NDA, the loop diagram which
eliminates the technifermions, and couples the $Z$,
one sees that the factor of $1/f^2$ is multiplied by
$\Lambda^2$, where $\Lambda$, the cutoff, is approximately
$4 \pi v$.  The two factors of $4 \pi$ eliminate the loop
suppression factor.
 In fact, the
$4 \pi$ suppression of the operators in eqs. (\ref{op}, 9) only comes
from replacing the factor of $1/f^2$ by
 $m_t/4 \pi v^3$.  If one were
now to consider a correction to a higher dimension
operator involving the gauge field, for example a
magnetic moment operator, the loop factor of $1/16 \pi^2$
cannot be fully compensated. It is essentially this
fact which keeps the ETC contribution to \bsg comparable
or suppressed relative
to that of the standard model, as we will now see.

Let us first consider the contribution to \bsg in a traditional
ETC scenario.  One might consider taking the ETC exchange which leads to the
bottom quark mass, integrating out a technifermion loop, and attaching a photon
to some part of the loop.
However, because the ETC gauge boson mass eigenstate exchanged couples to the
bottom quark mass eigenstate fields the magnetic moment operator induced
 is flavor diagonal. The
dominant (model independent) contribution to \bsg in fact comes from the same
ETC exchange which induced eq. (5), connecting
purely left-handed doublets. Integrating out a technifermion loop and attaching
the photon to the  technifermion line  now yields,
for example, the operator
\begin{equation}
 {\xi_3 \over (4 \pi)^2} ~{1 \over f^2}~
\overline{\psi}_L \partial\!\! \!\!\slash~
\sigma^{\mu \nu} \psi_L {e \over 2}
F_{\mu\nu},
\end{equation}
where $\psi_L$ is the  left-handed doublet containing the top quark and
$\xi_3$ contains order-one strong interaction and model dependence
 uncertainty. $e/2$ is the charge of the technifermion.
 The down component in $\psi$ is not exactly the bottom mass eigenstate,
 so (using the equations of motion) we can pull out the operator
\begin{equation}
 {m_t \over 4 \pi v}~ {m_b V_{ts} \over (4 \pi v)^2} \overline{b}_R
\sigma^{\mu \nu} s_L {e \over 2}
F_{\mu\nu}.
\end{equation}
There are also other contributions of comaparable magnitude.

 The standard model  induces an effective operator at one loop \cite{sbgam},
\begin{equation}
 A(m_t^2/M_W^2) m_b V_{ts}/(4 \pi v)^2~
\overline{b}_R \sigma_{\mu\nu} s_L~ e F^{\mu\nu},
\end{equation}
where $A$ is an order one function over the top mass range.
However, when QCD radiative corrections are included at the two loop
level in the standard model,
 there are large corrections to the one--loop result,
which increase the  prediction substantially \cite{qcd}.
The increased value comes from the mixing of  four-quark
operators into the operator with a photon. Clearly the nonstandard contribution
in traditional ETC is suppressed relative to the standard model by an amount of
order $m_t/(4 \pi v)$. We
therefore expect the rate for \bsg to agree with that from the standard model
at the $10\%$ level.

In techni-GIM models the bottom quark mass
is the sum of contributions from the exchange of several (nearly degenerate)
ETC gauge boson mass
eigenstates. Attaching a photon to the  ETC gauge bosons  and
integrating out the technifermions will now give a different linear combination
of contributions to the magnetic moment operator, permitting it to be
flavor-changing. More explicitly, the effective lagrangian will contain the
 higher dimension operator
\begin{equation}
{\xi_4 g_{ETC}^2 \over m_{ETC}^4}~
(\overline{d}_R \gamma^{\mu} D_R)~Y_d^{\dag}
Y_u Y_u^{\dag}~ (\overline{T}_L \gamma^{\nu} \psi_L) ~{e \over 6} F_{\mu \nu},
\end{equation}
arising from the exchange of (charge $1/6$)
 ETC gauge bosons with a photon attached.
Upon technifermion condensation and electroweak symmetry breaking this yields
\begin{equation}
{\xi_4 g_{ETC}^2 4 \pi v^3 \over 4 m_{ETC}^4}~
(\overline{b}_R ~Y_d^{\dag 33}
Y_u^{33} Y_u^{\dag 32}~ \sigma^{\mu \nu} s_L)~ {e \over 6} F_{\mu \nu}.
\end{equation}
With the estimate $Y_u^{23} \sim V_{ts}$, we get our final expression for
the effective operator,
\begin{equation}
{\xi_4 m_b V_{ts} \over 4 m_{ETC}^2} ~ \overline{b}_R \sigma_{\mu\nu} s_L~
  {e \over 6} F^{\mu\nu}.
\end{equation}

Now the ETC gauge boson mass should not be smaller than $4 \pi v$ in order to
keep the ETC physics separate from the strong technicolor dynamics.
 But the mass should also not
be much larger in
light of the large top quark mass.
Therefore we will estimate $m_{ETC} \sim 4 \pi v$ to get
\begin{equation}
{1 \over 24} ~{\xi_4 m_b V_{ts} \over (4 \pi v)^2} ~
\overline{b}_R \sigma_{\mu\nu} s_L~  e F^{\mu\nu}.
\end{equation}
The diagrams in which the photon is attached to the technifermion line rather
than the ETC gauge boson only make contributions to  dimension-six operators
analagous to eq. (18). Therefore their contributions to \bsg are much smaller
than eq. (24) despite the fact
that the technifermion charge is larger than the ETC gauge boson's.

We see that the
nonstandard contribution in techni-GIM models is probably numerically
suppressed, but (unlike traditional ETC models) not parametrically suppressed,
relative to the standard model. Still, the small nonstandard contribution
implies that these models should yield the standard model rate at the
 $10\%$ level.
It is possible that the techni-GIM mechanism is
incorporated in the context of strongly interacting dynamics beyond
technicolor itself, as suggested in the original ``CTSM" scenario.
In such a situation
we cannot trust the numerical suppression and
the nonstandard contribution may indeed be of the same order as the one-loop
standard model contribution. Thus a \bsg rate slightly larger than in the
 standard model could be accomodated.

\section{Conclusions}

We have studied the nonstandard contributions to the
 flavor-changing processes
$B_s \rightarrow \mu^+ \mu^-$ and $b \rightarrow s \gamma$
 in ETC theories, both
those with and without a techni-GIM mechanism. Within traditional ETC scenarios
without a GIM mechanism we obtained lower bounds on nonstandard contributions
to the rare decays which followed from the existence of
the minimal structure necessary for describing the ETC origins of the top
quark.
In the case of techni-GIM models, our estimates were based on
the existence
of approximately degenerate ETC gauge bosons but with large
flavor symmetry breaking to generate the top quark mass.
 The technicolor strong interaction matrix elements
we required were estimated using naive dimensional analysis. Thus our
predictions contain order one strong interaction uncertainties as well as order
one model dependence. And of course traditional ETC scenarios might contain
model--dependent contributions  larger than those we have estimated.

In both classes of models we found that   $b \rightarrow s \gamma$ generically
 occurs at essentially the standard model rate,
  although it is possible to accomodate a slightly larger rate in techni-GIM
  models if the ETC dynamics is strongly interacting.
 $B_s \rightarrow \mu^+ \mu^-$ occurs in both types of models mediated by a
 $Z$. If this is the only nonstandard contribution the rate is still likely to
 be an order of magnitude larger than in the standard model.
 However, the simplest  techni-GIM models also contain four-fermion
 interactions whose contributions raise the rate to two orders of magnitude
 above the standard model rate. The enhanced \bsm rates could be visible in the
 future.

\section*{Acknowledgements}

We thank Gordy Kane  and the 1993 Moriond Conference on Electroweak
Interactions
for motivating this work, Marjorie Shapiro for a useful conversation
 and Markus Luty for critical reading of the
manuscript. Lisa Randall thanks the CERN theory group for their hospitality.
 Raman Sundrum acknowledges support by the Director, Office
of Energy Research, Office of High Energy and Nuclear Physics, Division of High
Energy Physics of the U.S. Department of Energy under contract
DE-AC03-76SF00098 and by the National Science Foundation under grant
PHY90-21139.

\newpage

\section{Figure Caption}

Figure 1: $\Gamma(B_s \rightarrow \mu^+ \mu^-)/\Gamma(b \rightarrow c e^-
\overline{\nu}_e)$ as a function of $m_t$ in (a) technicolor models with only
the virtual $Z$ contribution, (b) technicolor with the direct four-fermion
contribution expected in the simplest techni-GIM models, (c) the standard
model. We have taken $f_B = 200$ MeV and all order one uncertainties to be
exactly one.


\newpage

\begin{thebibliography}{99}

\bibitem{bgam} E. Thorndike, CLEO Collab., Talk given at the 1993 Meeting of
the American Physical Society, Washington D.C., April 1993.

\bibitem{bmu} Proposal for an upgraded CDF detector, No. CDF/DOC/PUBLIC/1172
(1990).

\bibitem{2hig} S. Bertolini, F. Borzumati and A. Masiero, Nucl. Phys. B294
(1987) 321; T.G. Rizzo, Phys. Rev. D38 (1988) 820;
B. Grinstein and M. Wise, Phys. Lett. B201 (1988) 274; W.S. Hou and
R.S. Willey, Phys. Lett. B202 (1988) 591; T.D. Nguyen and G.C. Joshi, Phys.
Rev. D37 (1988) 3220; C. Q. Geng and J.N. Ng, Phys. Rev. D38 (1988) 2857; D.
Ciuchini, Mod. Phys. Lett. A4 (1989) 1945; J.L. Hewett, S. Nandi and T.G.
Rizzo, Phys. Rev. D39 (1989) 250;  B. Grinstein, R. Springer and M.
Wise, Nucl. Phys. B339 (1990) 269; V. Barger, J.L. Hewett and R.J.N. Phillips,
Phys. Rev. D41 (1990) 3421, and erratum;
S. Bertolini, F. Borzumati, A. Masiero and G. Ridolfi, Nucl. Phys.
B353 (1991) 591;
J.L. Hewett, Argonne preprint ANL-HEP-PR-92-110 (1992); V.
Barger, M.S. Berger and R.J.N. Phillips, Madison preprint MAD/PH/730 (1992);
T. Hayashi, M. Matsuda and M. Tanimoto, Kyoto preprint KU-01-93 (1993); R.
Barbieri and G.F. Giudice, CERN preprint CERN-TH.6830/93 (1993); M.A. Diaz,
Vanderbilt preprint VAND-TH-93-2 (1993).



\bibitem{sbmu} B. Grinstein, M.J. Savage and M.B. Wise,
Nucl. Phys. B319 (1989) 271; M.J. Savage, Phys. Lett. B266 (1991) 135;
A. Antaramian, L.J. Hall and A. Rasin, Phys. Rev. Lett. 69 (1992) 1871.

\bibitem{s} R. Renken and M. Peskin, Nucl. Phys. B211  (1983) 93;
B.W. Lynn, M.
Peskin and R.G. Stuart in `Physics at LEP', eds. J. Ellis and R. Peccei (1986);
 D. Kennedy and B.W. Lynn, Nucl. Phys. B322 (1989) 1;  M. Golden and L.
Randall,
Nucl. Phys. B361, (1991) 3; B. Holdom and J. Terning, Phys.
Lett. 247B  88 (1990); M. Peskin and T. Takeuchi, Phys. Rev. Lett. 65
964 (1990);
 A. Dobado, D. Espriu and M. Herrero, Phys. Lett. 255B  (1990)
405; W.J. Marciano and J.L. Rosner, Phys. Rev. Lett. 65 (1990) 2963;
 R. Johnson, B.L. Young and D.W. McKay, Phys. Rev. D43 (1991) R17.



\bibitem{gim} S. Glashow, J. Iliopoulos and L. Maiani, Phys. Rev. D2 (1970)
1285.

\bibitem{tgim} S. Dimopoulos, H. Georgi and S. Raby, Phys. Lett. B127 (1983)
101; R.S. Chivukula and H. Georgi, Phys. Lett. B188 (1987) 99;
R.S. Chivukula, H. Georgi and L. Randall, Nucl. Phys. B292 (1987) 93; H.
Georgi, Talk Presented at SCGT 90, Nagoya, Japan (1990), published in Nagoya
SCGT 90:155.

\bibitem{models} L. Randall, MIT preprint MIT-CTP\#2112 (1992).

\bibitem{model2} H. Georgi, Harvard preprint HUTP-92/A037 (1992).

\bibitem{fcnc} E. Eichten and K. Lane, Phys. Lett. B90 (1980) 125; S.
 Dimopoulos and J. Ellis, Nucl. Phys. B182 (1981) 505.


\bibitem{comp} T. Appelquist, T. Takeuchi, M. Einhorn and L.C.R. Wijewardhana,
Phys. Lett. B220 (1989) 223; T. Takeuchi, Phys. Rev. D40 (1989) 2697; V.A.
Miransky and K. Yamawaki, Mod. Phys. Lett. A4 (1989) 129; R.S. Chivukula, A.G.
Cohen and K. Lane, Nucl. Phys. B343 (1990) 554;
T. Appelquist, J. Terning and L.C.R. Wijewardhana, Phys. Rev.
D44 (1991).


\bibitem{fund} E.H. Simmons, Nucl. Phys. B312 (1989) 253;
S. Samuel, Nucl. Phys. B341 (1990) 513;
M. Dine, A. Kagan and S. Samuel, Phys. Lett. B243 (1990) 250; A. Kagan and S.
Samuel, in ``Quarks, Symmetries and Strings'', New York (1990).


\bibitem{walk} B. Holdom, Phys. Lett. 150B (1985) 301;
V.A. Miransky, Nuovo Cimento 90A
(1985) 149; T. Appelquist, D. Karabali
and
L.C.R. Wijewardhana, Phys. Rev. Lett. 57 (1986) 957;
 M. Bando, T. Morozumi, H. So
and
K. Yamawaki, Phys. Rev. Lett. 59 (1987) 389;
T. Appelquist and L.C.R. Wijewardhana, Phys. Rev. D36 (1987) 568.

\bibitem{nda} A. Manohar and H. Georgi, Nucl. Phys. B234 (1984) 189; H. Georgi
and L. Randall, Nucl. Phys. B276 (1986) 241.


\bibitem{zfc} L. Randall, Phys. Lett. B297 (1992) 309.

\bibitem{zbb} L. Randall and R.S. Chivukula, Phys. Lett. B202 (1988) 429.

\bibitem{zbb2} R.S. Chivukula, S.B. Selipsky and E.H. Simmons, Phys. Rev. Lett.
69 (1992) 575.

\bibitem{sbgam} T. Inami and C.S. Lim, Prog. Theor. Phys 65 (1981) 297.

\bibitem{qcd} B. Grinstein, R. Springer, M.B. Wise, Phys. Lett. B202 (1988)
138;
R. Grigjanis, P.J. O'Donnell, M. Sutherland  and H. Navelet, Phys. Lett.
  B224 (1989) 209, and B213 (1988) 355, erratum, B286, 413;
G. Cella, G. Curci, G. Ricciardi and A. Vicere, Phys. Lett. B248 (1990) 181.


\end{thebibliography}
\end{document}